\documentclass{PoS}

\title{Searches for BSM physics through CP violation at CDF.}

\ShortTitle{Searches for BSM physics through CP violation at CDF.}

\author{\speaker{Sabato Leo}, on behalf of the CDF collaboration\\
       University and INFN Pisa\\
       E-mail: \email{sabato.leo@pi.infn.it}\\}


\abstract{
The CDF experiment at the Tevatron $p\bar{p}$ collider has pioneered and established 
the role of hadron collisions in exploring flavor physics through a broad program that is now at its full maturity.
We report new results sensitive to physics
beyond the standard model, obtained using the whole CDF data set; 
including new bounds on the \Bs\ mixing phase and the decay width difference of \Bs\ mass-eigenstates and 
a measurement of the difference of CP asymmetries in $K^+K^-$ and $\pi^+\pi^-$ decays of $D^0$ mesons. 
We also present a new measurement of the $\Bs \rightarrow D^{(*)+}_s D^{(*)-}_s$ branching ratio using 6.8 fb$^{-1}$
of data and search for CP violation in $D^0 \rightarrow K^0_s \pi^+ \pi^-$ decays in 6.0 fb$^{-1}$ of data.
}

\FullConference{The XIth International Conference on Heavy Quarks and Leptons \\
                11-15 June, 2012\\
                Prague,The Czech Republic}

\def\Journal#1#2#3#4{{#1} {\bf #2}, #3 (#4)}

\def\NIMA{{\em Nucl. Instrum. Methods} A}

\def\PLB{{\em Phys. Lett.}  B}
\def\PRL{\em Phys. Rev. Lett.}
\def\PRD{{\em Phys. Rev.} D}

\def\dpipi{\ensuremath{D^0\to \pi^+\pi^-}}
\def\dKK{\ensuremath{D^0\to K^+K^-}}
\def\dAcp{\ensuremath{\Delta A_{\rm{CP}}}}
\def\betas{\ensuremath{\beta_s}}

\def\DGs{\ensuremath{\Delta\Gamma_s}}
\def\Bs{\ensuremath{B^0_s}}
\def\B0{\ensuremath{B^0}}
\def\bBs{\ensuremath{\bar{B}^0_s}}

\def\jpsiphi{\ensuremath{\Bs \myto \Jpsi\mbox{ }\phi}}

\def\B0    {\ensuremath{B^0}}

\def\Bs    {\ensuremath{B^0_s}}
\def\bBs  {\ensuremath{\bar{B}^0_s}}

\def\Jpsi{\ensuremath{J\!/\!\psi}}
\newcommand{\myto}{\kern -0.3em\to\kern -0.2em}
\def\BsJpsiPhi{\ensuremath{\Bs \myto \Jpsi \mbox{ }\phi}}
\def\betas     {\ensuremath{\beta_s}}
\def\DGs       {\ensuremath{\Delta\Gamma_s}}

\def\taus   {\ensuremath{\tau_{s}}}

\newcommand{\TeV}{\ensuremath{\matqq	hrm{Te\kern -0.1em V}}}
\newcommand{\TeVc}{\ensuremath{\mathrm{Te\kern -0.1em V\!/}c}}
\newcommand{\TeVcc}{\ensuremath{\mathrm{Te\kern -0.1em V\!/}c^2}}
\newcommand{\GeV}{\ensuremath{\mathrm{Ge\kern -0.1em V}}}
\newcommand{\GeVc}{\ensuremath{\mathrm{Ge\kern -0.1em V\!/}c}}
\newcommand{\GeVcc}{\ensuremath{\mathrm{Ge\kern -0.1em V\!/}c^2}}
\newcommand{\MeV}{\ensuremath{\mathrm{Me\kern -0.1em V}}}
\newcommand{\MeVc}{\ensuremath{\mathrm{Me\kern -0.1em V\!/}c}}
\newcommand{\MeVcc}{\ensuremath{\mathrm{Me\kern -0.1em V\!/}c^2}}

\newcommand{\cdfii}{CDF\,II~}
\newcommand{\ppbar}{\ensuremath{p\overline{p}}}

\newcommand{\KpKm}{\ensuremath{K^+K^-}}

\newcommand{\CP}{\ensuremath{C\!P}}

\renewcommand{\epsilon}{\varepsilon}
\renewcommand{\theta}{\vartheta}

\begin{document}

\section{Introduction}

Flavor physics of quarks provides one of the most promising probes for indirect signs of new particles or 
interactions beyond the standard model (SM).
Strange and charm meson dynamics in particular offers rich opportunities since its
experimental exploration has not reached in extension and precision that has been achieved in processes involving charged and neutral kaons and 
bottom mesons.\par
The \Bs\ oscillations are explained in terms of second-order weak processes involving the CKM matrix element $V_{ts}$.
A broad class of generic extensions of the SM is expected to affect the mixing amplitude, modifying the mixing ``intensity''--that is 
the oscillation frequency--and the phase, $\betas = arg[-(V_{ts}V^*_{tb})/(V_{cs}V^*_{cb})]$. A non-SM enhancement of \betas\  would also decrease the size of 
the decay-width difference \DGs\ between the light and heavy mass eigenstates of the \Bs\ meson~\cite{Faller:2008gt}. 
While the oscillation intensity has been measured precisely~\cite{Abulencia:2006ze}, only loose constraints on the phase and width difference were available until recently. 
The most effective determination of \betas\ and \DGs\ is achieved through the analysis of the time evolution of ``flavor--tagged'' \BsJpsiPhi\ decays.
The first such analysis was performed by CDF in 2008~\cite{sin2betas-early}. 
D0 followed with a similar measurement~\cite{Abazov:2011ry} soon after CDF. In 2010 the combination of CDF and D0 results suggested a mild deviation from the SM expectation.
However, updated measurements ~\cite{Abazov:2011ry,CDF:2011af,LHCb:2011aa} showed increased 
consistency with the SM, calling for additional experimental information to clarify the picture. Here we report the latest CDF update using the final data set of 10 fb$^{-1}$.
We also report a measurement of the $\Bs \rightarrow D^{(*)+}_s D^{(*)-}_s$ branching ratios using 6.8 fb$^{-1}$ of CDF data, which provides information on the ``width--difference''.\par
The observation of a sizable $D^0$--$\bar{D}^0$ mixing~\cite{babarDmix,BelleDmix,CDFDmix} has raised an increasing interest in charm dynamics, 
where CP violation may play an important role. Within the SM, ``CP--violating'' effects are predicted to be
small since the charm transitions are described, to an excellent approximation, by physics
of the first two generations of quarks. The size of CP violation expected from the Cabibbo-Kobayashi-Maskawa hierarchy is $\mathcal{O}(10^{-3})$ or less.
However both the $D^0$--$\bar{D}^0$ mixing amplitude and the SM-suppressed penguin amplitude
can be greatly enhanced by new dynamics, which can also increase the size of the CP violation. 

Among the most sensitive probes of physics beyond the SM are the tree-dominated decays \dpipi\ and  \dKK .
Any measured direct asymmetry significantly larger than 1\%~would be indication of NP.
Last year, using 5.9 fb$^{-1}$ of data, CDF produced the world's most precise measurements of 
the CP asymmetries $A_{\rm{CP}}(KK)=(-0.24\pm0.22\pm0.09)\%$ and  $A_{\rm{CP}}(\pi\pi)=(0.22\pm0.24\pm0.11)\%$~\cite{Acp_cdf}.
In spite of the hadronic uncertainties, there is some consensus that direct CP asymmetries of \dKK\ and of \dpipi\ should be of opposite sign. 
Therefore, a measurement of the difference between asymmetries of those decays
 is maximally sensitive to detect direct CP violation. 
Indeed, the LHCb collaboration reported recently the first evidence of CP violation in charm 
measuring $\dAcp =A_{\rm{CP}}(KK)-A_{\rm{CP}}(\pi\pi)= (-0.82 \pm 0.21 \pm 0.11)\%$~\cite{dAcp_lhcb}.
An independent measurement is crucial to establish the effect, 
 and the 10 fb$^{-1}$ sample of hadronic $D$ decays collected by CDF is the only one currently available 
to attain sufficient precision. 

Another suitable channel to search for CP violation in charm is the ``three--body'' decay $D^0 \rightarrow K^0_s \pi^+ \pi^-$ and its resonant substructure.
%
%
In sec.\ref{sec_D0Kspipi} we report the first such measurement in hadron collisions using 6 fb$^{-1}$ of data.

\section{The CDF II detector}
The collider detector at fermilab is a multipurpose experiment designed to study $\sqrt{s}=$1.96TeV \ppbar\ collisions produced 
by the Tevatron collider in 2001--2011.
Among the components and capabilities of the CDF II detector~\cite{detector}, the tracking system is the one most relevant
to these analyses. It lies within a uniform, axial magnetic field of 1.4 T strength. The inner tracking volume up to a
radius of 28 cm\,is composed of 6-7 layers of ``double--sided'' silicon ``micro--strip'' detectors~\cite{Hill}. An additional layer of ``single--sided'' 
silicon is mounted directly on the ``beam-pipe'' at a radius of 1.5 cm, allowing excellent resolution on the impact
parameter $d_0$, defined as the distance of closest approach of a reconstructe ``charged--particle'' track to the interaction point in the plane transverse
to the beam line. The silicon detector provides a vertex resolution of approximately 15 $\mu m$ in the transverse and
70 $\mu m$ in the longitudinal direction. The remainder of the tracking volume from a radius of 40 to 137 cm is occupied
by an ``open--cell'' drift chamber (COT)~\cite{cot}, providing a transverse momentum resolution of $\sigma_{p_{T}}/p_{T} \approx 0.07\% p_T$ ($p_T$ in GeV/$c$).
Hadron identification, which is crucial for distinguishing slow kaons and protons from pions and muons, is achieved by a likelihood
combination of information from a time-of-flight system~\cite{tof} and ionization energy loss in the COT. This offers
about 1.5$\sigma$ separation between kaons and pions.
A ``three--level'' trigger system is used for the online event selection. At level 1 the most important device for the described analyses
is the extremely fast tracker (XFT)~\cite{fast_tracker}. It identifies charged particles using information from the COT
and measures their transverse momenta and azimuthal angles around the beam direction. The basic requirement
at level 1 is two charged particles with transverse momentum greater than 1.5 GeV/$c$. 
At level 2 and 3, different requirements are imposed for the different analyses: two ``oppositely--charged'' particles reconstructed in 
the ``drift--chamber'' matched to ``muon--chamber'' track segments with a dimuon mass consistent with the $J/\psi$ mass (``low--$p_T$'' dimuon trigger) for the analysis 
in sec.~\ref{sec:jpsiphi}; for the remaining analyses, the trigger uses XFT tracks combined with tracks reconstructed in the silicon tracker,
thus allowing the precise measurement of impact parameters of tracks~\cite{punzi_ristori}. Impact parameters are required to be 
between 0.1 and 1 mm and to be consistent with coming from a common vertex displaced from the interaction point by at least 
100 $\mu$m in the plane transverse to the beam line. 

\section{Measurement of the \jpsiphi\ time-evolution in the final CDF Run II data set}\label{sec:jpsiphi}

\begin{figure}[hbtp]
\begin{center}
\includegraphics[width=0.45\textwidth,height=0.45\textwidth]{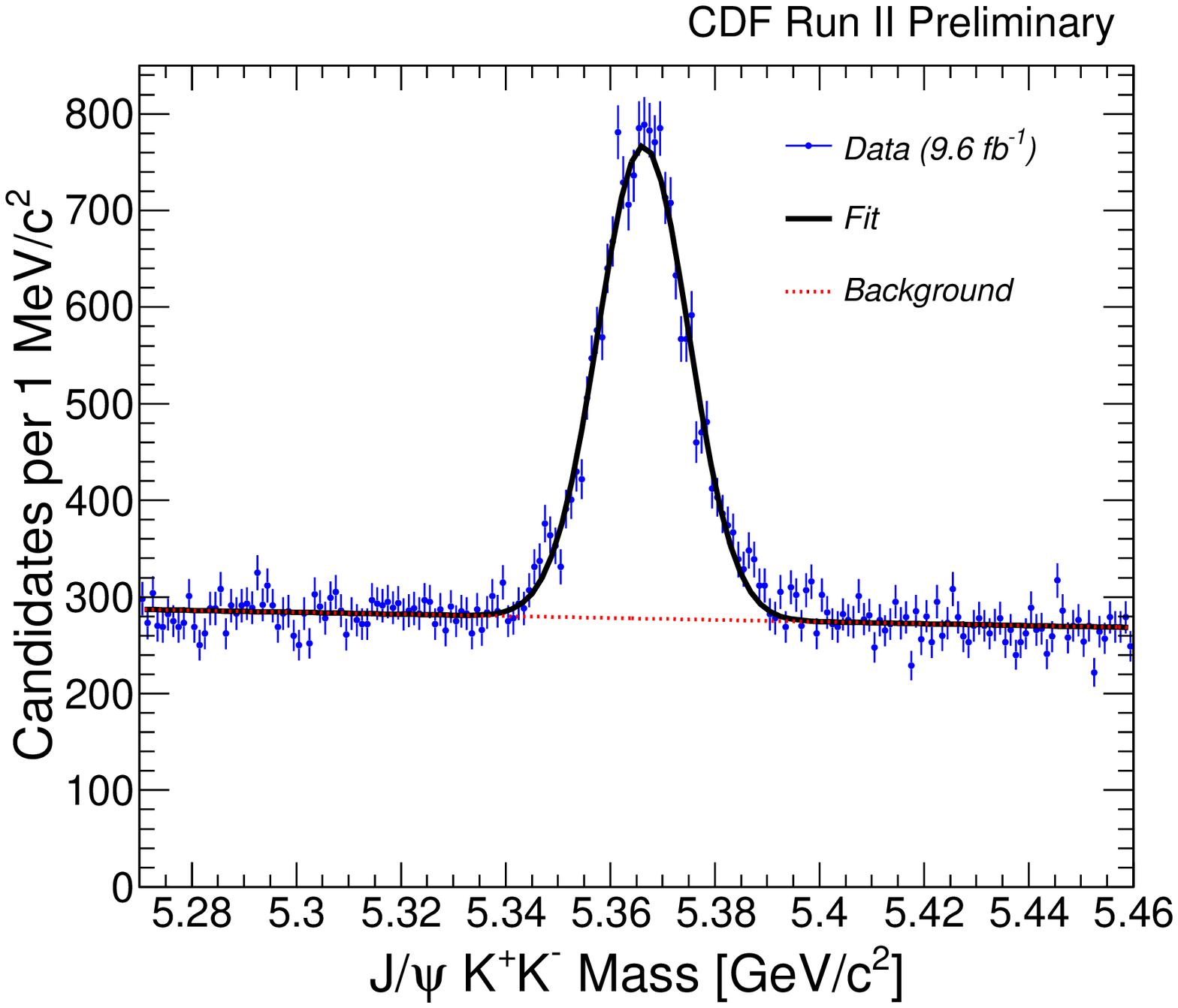}
\includegraphics[width=0.45\textwidth,height=0.45\textwidth]{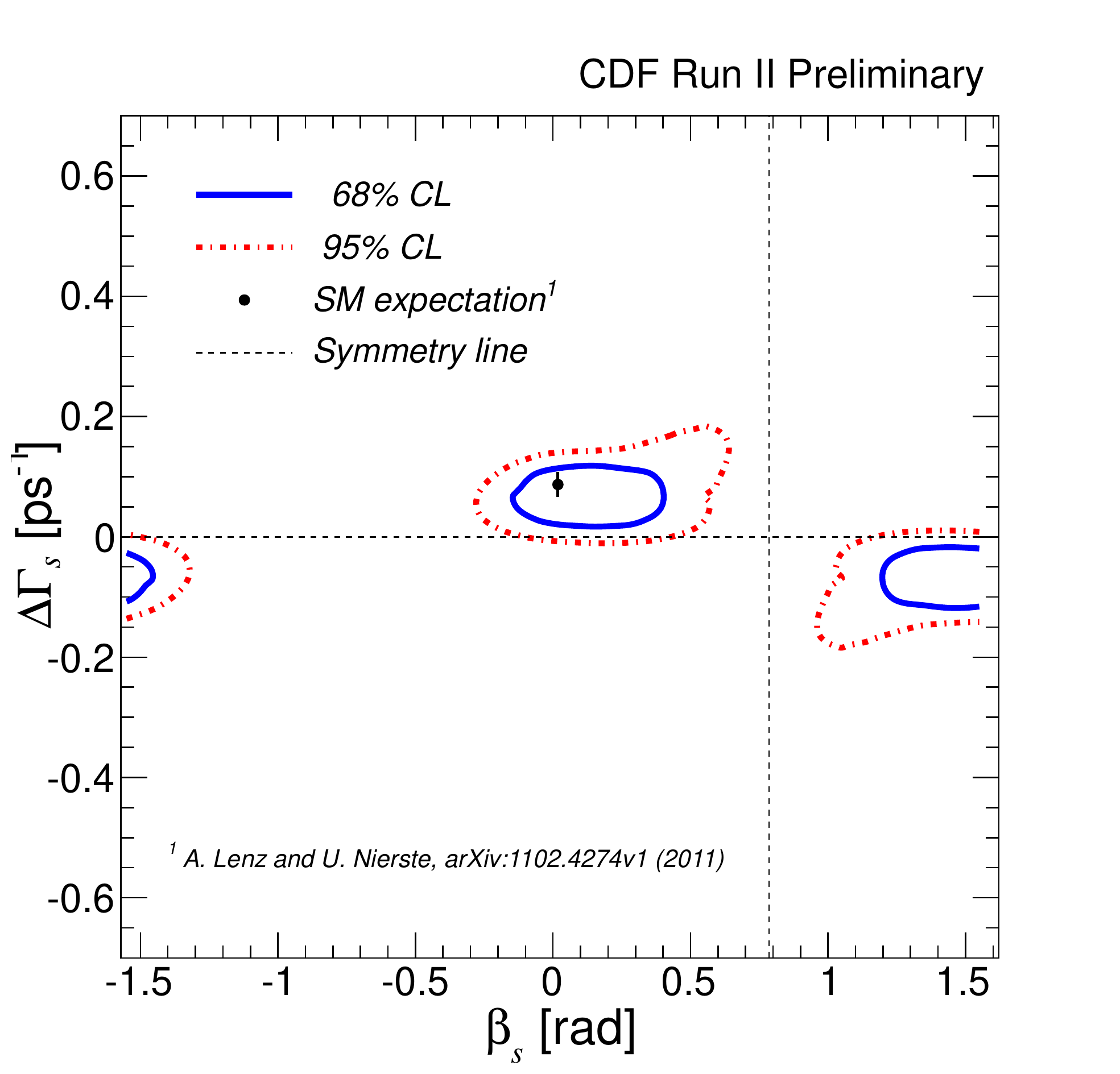}
\end{center}
\caption{\textbf{Left:} Distribution of $\Jpsi K^+K^-$ mass with fit projection overlaid. \textbf{Right:} Confidence regions at the 68\% 
and 95\% CL in the (\betas, \DGs) plane.}
\label{fig:mass}
\label{fig:contour}
\end{figure}

The \jpsiphi\ decays are fully reconstructed using four tracks originating from a common displaced vertex, 
two matched to muon pairs consistent with a \Jpsi\ decay ($3.04< m_{\mu\mu}< 3.14$ \GeVcc), and two consistent with a $\phi \myto \KpKm$ decay ($1.009< m_{KK}< 1.028$ \GeVcc).
The dimuon mass constraint to the known \Jpsi\ mass, combined with the good $p_T$ resolution, yield a mass resolution of the signals of about 9 MeV/$c^2$.
The $\Jpsi K^+K^-$ mass distribution (fig.~\ref{fig:mass}, left), shows a signal of approximately 11~000 decays, overlapping a constant background dominated by the prompt combinatorial component and smaller contributions from mis-reconstructed $B$ decays.
The analysis relies on a joint fit to the time evolution of \Bs\ mesons that resolves the fast oscillations by exploiting 
the 90~fs time resolution of the CDF silicon detector for these final states.
Because the $B^0_s$ meson has spin zero and \Jpsi\ and $\phi$ have spin one,   
the \BsJpsiPhi\ decay involves three independent amplitudes, each corresponding to one possible angular momentum state 
of the $\Jpsi \phi$ system, which is also a \CP-odd or \CP-even eigenstate.
To enhance the sensitivity to \betas, the time-evolution of the three decay amplitudes is fit independently by exploiting differences 
in the distribution of the kaon' and muon' decay angles.
Sensitivity to \betas\ can be further enhanced by accounting for the difference in the time evolution of initially produced \Bs\ and \bBs\ mesons.
The flavor of the meson at the time of production is inferred by two independent classes of algorithms:
the opposite-side flavor tag (OST) and the same-side kaon tag (SSKT)~\cite{Abulencia:2006ze}.
OST algorithms infer the initial flavor of the meson candidate from the decay products of the $b$ hadron produced by the other $b$ quark in the event;
SSKT algorithms deduce the production flavor exploiting the ``charge--flavor'' correlation of the neighboring kaons produced in the fragmentation process.
The OST performance has been determined with 82~000 $B^\pm \myto \Jpsi(\myto \mu^+\mu^-) K^\pm$ decays fully reconstructed in the same 
sample as the signal. 
We found an efficiency of $\epsilon_{\rm{OST}} = (92.8\pm0.1)\%$, an observed averaged dilution, $D_{\rm{OST}} = 1 -2 w_{\rm{OST}}$, equal to $(12.3 \pm 0.4)\%$ and a resulting effective tagging power of $\epsilon_{\rm{OST}}D^2_{\rm{OST}} = (1.39 \pm 0.05)\%$. 
The SSKT algorithms tag a smaller fraction of candidates with better precision. 
Its performance has been previously determined~\cite{CDF:2011af} to be $\epsilon_{\rm{SSKT}} = (52.2 \pm 0.7)\%$, $D_{\rm{SSKT}}=(21.8\pm 0.3)\%$ 
and $\epsilon_{\rm{SSKT}}D^2_{\rm{SSKT}}=(3.2 \pm 1.4)\%$.
Since the SSKT algorithm has been calibrated for early data only, we conservatively restrict its use to the events collected in that period. 
Simulation shows that this results in a modest degradation in \betas\ resolution.\par
The unbinned maximum likelihood joint fit uses 9 observables from each event to determine 32 parameters including $\beta_s$ and $\Delta\Gamma$,  other physics parameters (\Bs\ lifetime, decay amplitudes at $t=0$ and phases, etc),  and several other (``nuisance") parameters (experimental scale factors, etc.).
If \betas\ is fixed to its SM value, the fit shows unbiased estimates and Gaussian uncertainties for \DGs\ and \taus. 
We found $\DGs =  0.068 \pm 0.026(stat) \pm0.007(syst)$ ps$^{-1}$, and mean \Bs\ lifetime,  $\tau_s = 1.528 \pm 0.019(stat) \pm0.009(syst)$ ps.
Systematic uncertainties include mismodeling of the signal mass model, lifetime resolution,  acceptance description, and angular distribution of the background;
a $\mathcal{O}(2\%)$ contamination by $\B0 \myto \Jpsi K^*(892)^0$ decays misreconstructed as \BsJpsiPhi\ decays; and the silicon detector misalignment.
These results are among the most precise from a single experiment.
If \betas\ is free to float in the fit, tests in statistical trials show that the maximum likelihood estimate 
is biased for the parameters of interest, and the biases depend on the true values 
of the parameters. Hence, we determine confidence regions in the \betas\ and $(\betas,\DGs)$ spaces (fig.~\ref{fig:contour}, right), by using a profile-likelihood ratio statistic as a $\chi^2$ variable and considering all other likelihood variables as nuisance parameters.
Confidence regions are corrected for ``non--Gaussian'' tails and systematic uncertainties to ensure nominal coverage.
By treating \DGs\ as a nuisance parameter, we also obtain $\beta_s\in[-\pi/2,-1.51]\cup[-0.06,0.30]\cup [1.26,\pi/2]$ rad at the 68\% CL, and $\betas \in [-\pi/2, -1.36] \cup [-0.21,0.53] \cup [1.04,\pi/2]$ rad at the 95\% CL.
The fit also includes the CP-odd component that can originate by non resonant $K^+K^-$ pair or by the $f_0(980)$ decays. 
The resulting S-wave decay amplitude is found to be negligible.
All results are consistent with the SM expectation and with determinations of the same quantities from other experiments~\cite{Abazov:2011ry,LHCb:2011aa,Aaij:2012eq}.

\section{Measurement of $\Bs \rightarrow D^{(*)+}_s D^{(*)-}_s$ branching ratios}\label{sec:branching_ratios}

\begin{figure}[hbtp]
\begin{center}
\includegraphics[width=0.45\textwidth]{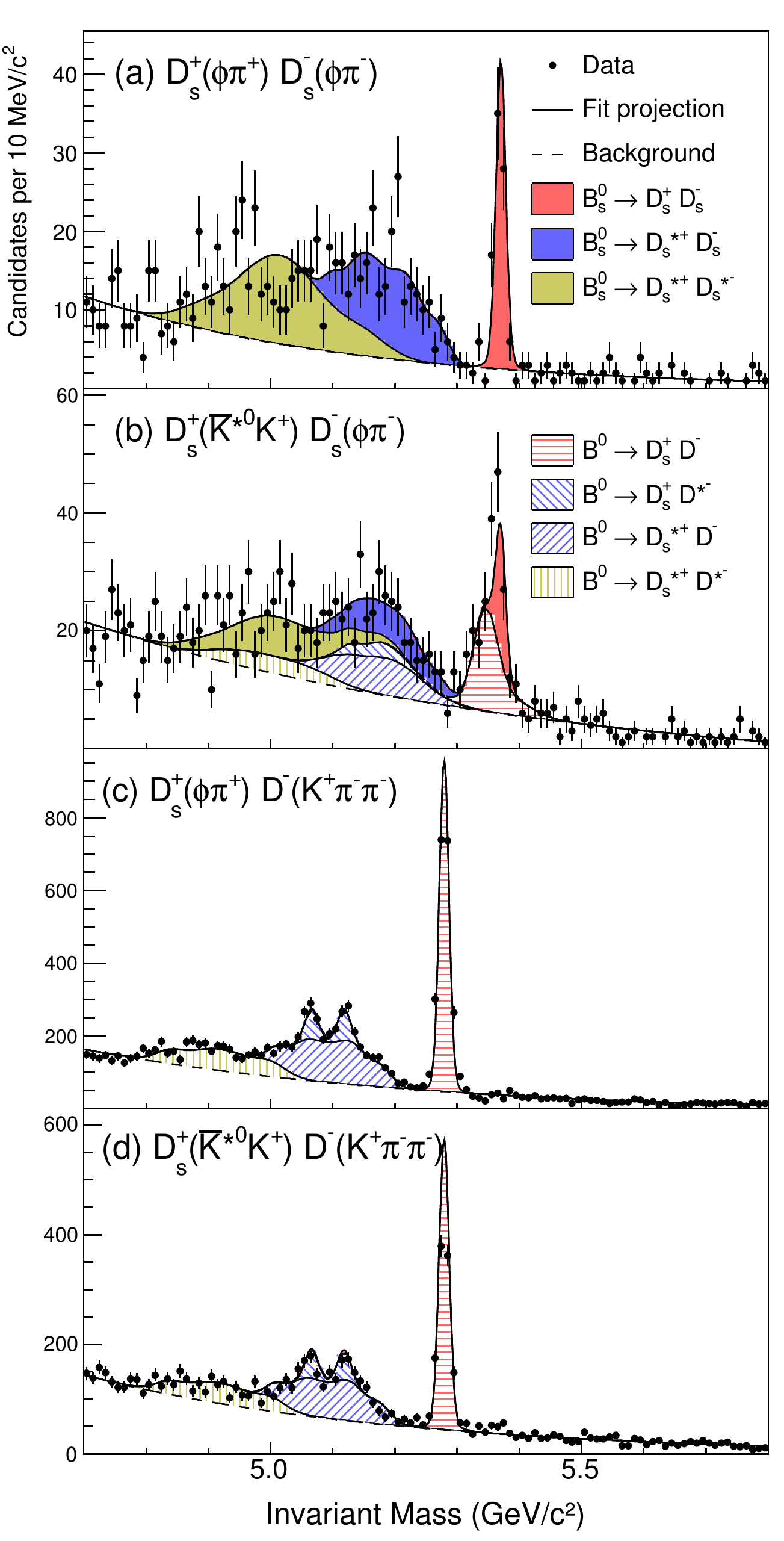}
\end{center}
\caption{Invariant mass distribution of $\Bs\ \rightarrow D^+_s(\phi \pi^+)D^-_s(\phi \pi^-)$,
$\Bs\ \rightarrow D^+_s(\bar{K}^{*0} K^+)D^-_s(\phi \pi^-)$, 
$B^0 \rightarrow D^+_s(\phi \pi^+)D^-(K^+ \pi^- \pi^-)$ and $B^0 \rightarrow D^+_s(\bar{K}^{*0} K^+)D^-(K^+ \pi^- \pi^-)$ candidates with 
the simultaneous fit projection overlaid. The broader structures stem from decays where the photon or $\pi^0$ from 
the $D^{*+}_{(s)}$ decay is not reconstructed.}
\label{fig:dsds}
\end{figure}

A measurement of \Bs\ production rate times the $\Bs \rightarrow D^{(*)+}_s D^{(*)-}_s$ branching ratio relative to the normalization 
mode $B^0 \rightarrow D^{+}_s D^{-}$ is perfomed using a data sample corresponding to an integrated luminosity of 6.8 fb$^{-1}$ 
recorded by the displaced track trigger~\cite{BsDsDs_ref}.
$D^+_s \rightarrow K^+ K^- \pi^+$ and $D^+ \rightarrow K^- \pi^+ \pi^+$ decays are reconstructed from combinations of three tracks 
with appropriate charge and mass hypothesis assignments, fitted to a common vertex and then combined into another vertex to 
form \Bs\ candidates. For the first time in this channel we 
exploit the Dalitz structure of the intermediate states ``three--body'' decays is exploited for an accurate evaluation  of acceptances and efficiencies.
The relative branching fractions are determined in a simultaneous maximum likelihood fit to the signal, $(\phi \pi^+)(\phi \pi^-)$ and 
$(\bar{K}^{*0} K^+)(\phi \pi^-)$, and background, $(\phi \pi^+)(K^+ \pi^- \pi^-)$ and $(\bar{K}^{*0} K^+)(K^+ \pi^- \pi^-)$, unbinned mass 
distributions. The components of the fit functions for each mass distribution are fully and partially reconstructed signals, reflections, 
and background.
Figure \ref{fig:dsds} shows the projections of the fit overlaid to data. The statistical significance of each signal exceeds 
10$\sigma$.
Using measured values of production and relative branching fractions, the following absolute
branching fractions are derived:
\begin{eqnarray*}
\mathcal{B}(B_s \rightarrow D^+_s D^-_s)             &=& (0.49 \pm 0.006 \pm 0.05 \pm 0.08)\%, \nonumber \\ 
\mathcal{B}(B_s \rightarrow D^{*\pm}_s D^{\mp}_s)    &=& (1.12 \pm 0.12  \pm 0.09 \pm 0.19)\%, \nonumber \\
\mathcal{B}(B_s \rightarrow D^{*+}_s D^{*-}_s)       &=& (1.79 \pm 0.19  \pm 0.27 \pm 0.29)\%, \nonumber \\
\mathcal{B}(B_s \rightarrow D^{(*)+}_s D^{(*-)}_s)   &=& (3.38 \pm 0.25  \pm 0.30 \pm 0.56)\%, \nonumber \\
\end{eqnarray*}

in which statistical, systematic and normalization uncertainties are reported. 
These results are the most precise to date from a single experiment and can provide information related to the ``decay--width'' difference \DGs .

\section{Measurement of CP violation in charm decays in the final CDF Run II data set}
\begin{figure}
\begin{center}
\includegraphics[width=0.85\textwidth]{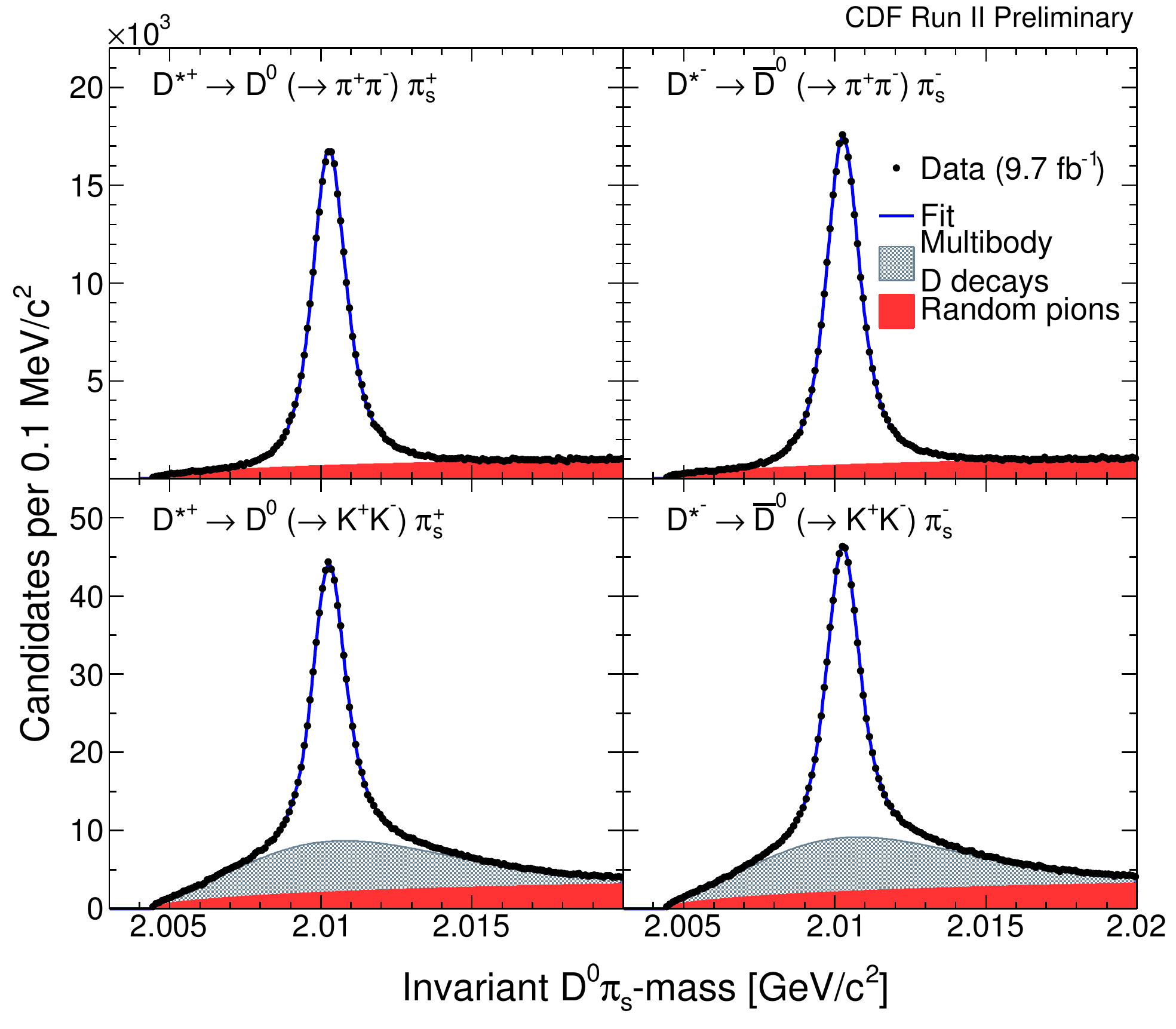} \hspace{10 mm}
\end{center}
\caption{Projections of the combined fit on data for tagged $D^0 \rightarrow K^+ K^-$ (top) and $D^0 \rightarrow \pi^+ \pi^-$ (bottom) decays. 
Charm decays on the left and anticharm on the right.}
\label{fig:charm_asymmetries}
\end{figure}

CDF previously measured CP violation in \dKK\ and \dpipi\ decays.  
In a larger data sample, we have measured the difference between those asymmetries with greater precision. 
The analysis follows closely the measurement of individual asymmetries~\cite{dAcp_cdf}.
The flavor of the $D^0$ meson is tagged from the charge of the soft pion in the 
strong $D^{\star +} \to D^0 \pi^+$ decay. Since $D^{\star +}$ and $D^{\star -}$ mesons are produced 
in equal number in $p\bar{p}$ interactions, any asymmetry between 
the number of $D^0$ and $\bar{D}^0$ decays is due to either CP violation or instrumental effects.
The latter can be induced only by the difference in reconstruction
efficiency between positive and negative soft pions. Provided that the relevant kinematic distributions 
are equalized in the two decay channels, the instrumental asymmetry cancels 
to an excellent level of accuracy in the difference between the observed asymmetries between signal yields.
Such cancellation allows an increase in sensitivity on \dAcp\ by loosening some selection criteria with respect to 
the measurement of individual asymmetries thus doubling the signal yields. 
The offline selection follows the standard trigger selection with some basic additional requirements on track and vertex quality.  
The numbers of $D^0$ and $\bar{D}^0$ decays are determined with a simultaneous fit to the $D^0\pi$-mass
distribution of positive and negative $D^\star$ decays. 
About 1.21 times $10^6$ \dKK\ decays, fig.~\ref{fig:charm_asymmetries} bottom, and 550 times $10^3$ \dpipi\ decays, fig.~\ref{fig:charm_asymmetries} top, 
are reconstructed, yielding the following 
observed asymmetries between signal yields, 
$A_{\rm{raw}}(KK)=(-2.33\pm0.14)\%$ and $A_{\rm{raw}}(\pi\pi)=(-1.71\pm0.15)\%$.
Residual systematic uncertainties 
total 0.10\% and are driven by differences 
between $D^*$ mass distributions associated with charm and anticharm decays. 
The final result is $\dAcp=(-0.62\pm0.21\pm0.10)\%$, 
which is 2.7$\sigma$ different from zero~\cite{arxiv_angelo}.
This provides strong indication of CP violation in CDF charm data, 
supporting the LHCb earlier evidence with the same resolution. 
 The combination of CDF, LHCb, and $B$-factory measurements  
deviates by approximately 3.8$\sigma$ from the no CP violation point.

\section{Search for CP violation in $D^0 \rightarrow K^0_s \pi^+ \pi^-$ decays}\label{sec_D0Kspipi}

We exploit a large sample of $D^{*+}$ decays with the $D^0$ production flavor determined by the charge of the soft pion in $D^{*+} \rightarrow D^0 \pi^+ $ decays. 
The offline reconstruction of candidates starts with refitting tracks using the pion mass hypothesis. Two oppositely charged
tracks are combined to form a $K^0_S$ candidate. To construct $D^0$ candidates, each $K^0_S$ candidate is then combined with
all possible ``oppositely--charged track--pairs'' from the remaining tracks in the event. Finally, the $D^{* +}$ candidates
are obtained by combining each $D^0$ candidate with one of the still remaining tracks in the event. The tracks forming
the $K^0_S$ , $D^0$ , and $D^{* +}$ candidates are subjected to kinematic fits that constrain them to originate from
common vertices. Standard quality requirements on tracks and vertices are used
to ensure ``well--measured'' masses and ``decay--positions''.
An appropriately trained neural network 
classifer contributes the final discrimination between signal and background.

\begin{figure}
\begin{center}
\includegraphics[width=0.65\textwidth]{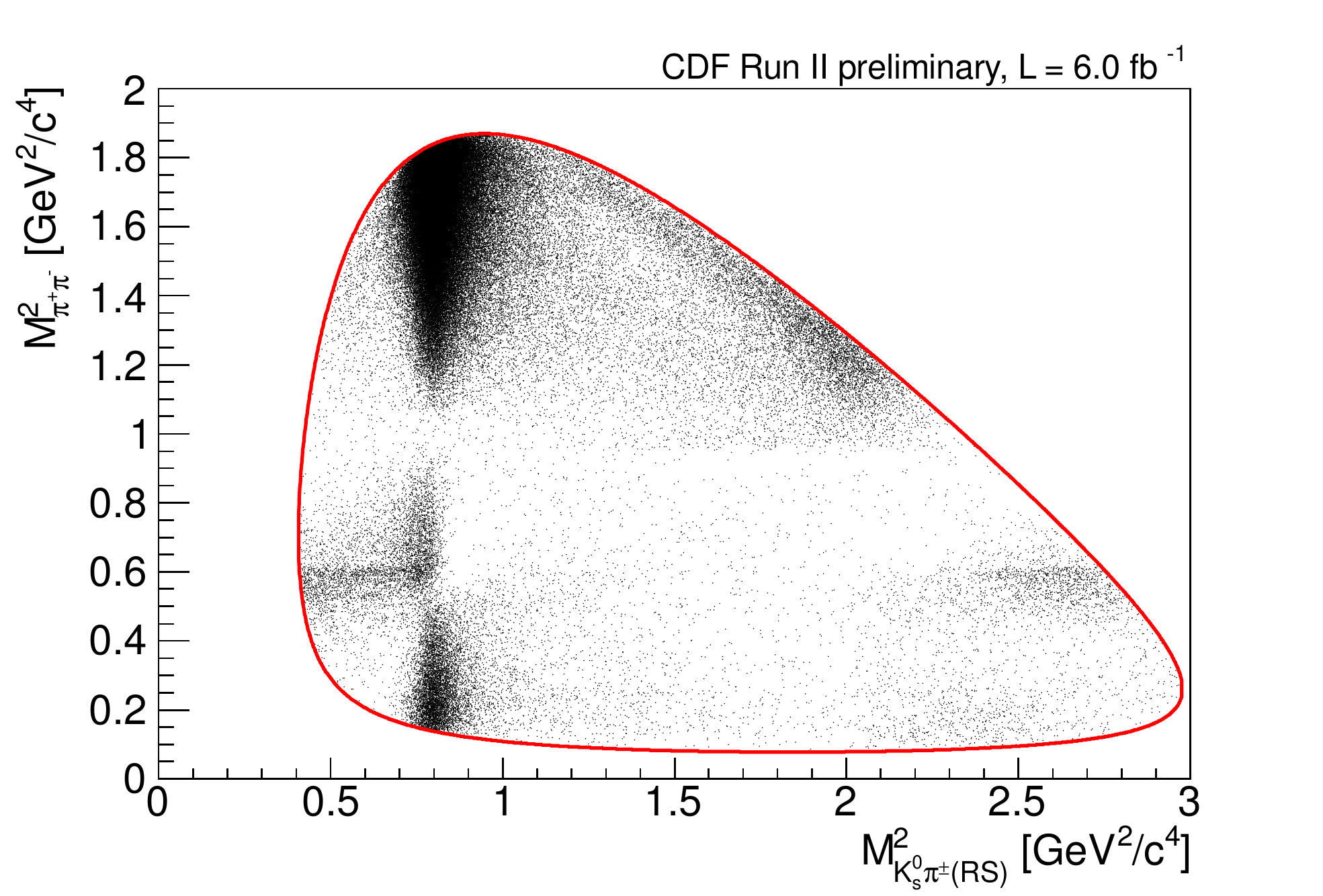} \hspace{10 mm}
\end{center}
\caption{Dalitz plot of the decay $D^0 \rightarrow K^0_s \pi^+ \pi^-$ , where the squared invariant masses of the ``two--body'' combinations 
$K^0_S \pi^-$ and $\pi^+ \pi^-$ are chosen as kinematic quantities. The red line indicates the kinematic boundaries.
}
\label{fig:dskpipi}
\end{figure}

The resonant substructure of a three-body decay is described using the Dalitz plot method~\cite{Dalitz_Mag}. 
The Dalitz plot of the considered decay $D^0 \rightarrow K^0_s \pi^+ \pi^-$, composed of all selected candidates, is shown in fig. \ref{fig:dskpipi}. 
Three types of intermediate resonances contribute:
Cabibbo allowed, doubly Cabibbo suppressed, and CP eigenstates. The dominant decay mode is the Cabibbo allowed
$D^0 \rightarrow K^*(892)^- \pi^+$ which amounts to about 60\% of the total branching fraction. The second largest contribution is
from the intermediate CP eigenstate $K^0_S \rho(770)$, which is color suppressed compared to $K^∗(892)^- \pi^+$.
A binned maximum likelihood fit to the ``two--dimensional'' Dalitz plot distribution with bin widths of 0.025 GeV$^2$/c$^4$
in both dimensions is performed to determine the contributions of the different intermediate resonances.
The isobar model is used to describe the form of the matrix element implemented in the likelihood.
The results of the simultaneously fit to $D^0$ and $\bar{D}^0$ Dalitz plots for the CP violating amplitudes and phases are reported in Ref.~\cite{CDF_DSKSpipi_results}.
The overall integrated CP asymmetry is found to be $A_{CP} = (-0.05 \pm 0.57 \pm 0.54)\%$ with a large improvement over previous results~\cite{CLEO_results}. All the CP violating quantities are found to be consistent with zero.
Following the so-called Miranda procedure~\cite{Bediaga}, we also performed a ``model--independent'' search for CP violation in 
the Dalitz plot distribution of the decay $D^0 \rightarrow K^0_s \pi^+ \pi^-$ by comparing the binned, normalized Dalitz
plots for $D^0$ and $\bar{D}^0$. No assumptions about the resonant substructure of the decay
are used. This approach confirms that ``no--CP--violation`` between the $D^0$ and $\bar{D}^0$ Dalitz plots is present.

\section{Summary}

The results of a few recent updates on flagship flavor measurements that use the complete \cdfii data set or a large fraction of it have been reported.
Improved bounds on the \Bs\ mixing phase and decay width difference of \Bs\ mass-eigenstates are found to be consistent with the standard model. 
We also presented the most precise measurement of $\Bs \rightarrow D^{(*)+}_s D^{(*)-}_s$ branching ratio.
A measurement of the difference of CP asymmetries in $K^+K^-$ and $\pi^+\pi^-$ decays of $D^0$ meson shows significant
discrepancy from zero confirming and supporting similar results from LHCb. Its interpretation as due to physics within or beyond the SM is still under debate.
We finally presented a null search for CP violation in $D^0 \rightarrow K^0_s \pi^+ \pi^-$ decays in 6.0 fb$^{-1}$ of data.\par
These measurements are a sampling from an intensive flavor physics program developed at the Tevatron in the last decade that is now at its full maturity.
The chief achievement has been to extend and deepen the study of the \Bs\ mesons dynamics, which was only marginally explored before.
Stringent constraints on the presence of NP in \Bs\ mixing has been imposed by measuring with a good precision 
both its intensity and its associated phase.
In addition, favorable production cross sections together with the CP-symmetric nature of the \ppbar\ collisions allowed exploration of
CPV in the charm sector with ``world--leading'' precision.
Another important heritage has been to show 
that flavor physics is in fact possible also at hadron colliders and provides information competitive and complementary to the information from dedicated experiments. 
Many trigger reconstruction, and data analysis techniques developed at CDF are now used as standard 
by the next generation of flavor experiments at hadron colliders.

\end{document}